\documentclass[conference]{IEEEtran}
\IEEEoverridecommandlockouts
\usepackage{subcaption}
\usepackage{cite}
\usepackage{amsmath,amssymb,amsfonts}
\usepackage{algorithmic}
\usepackage{booktabs}
\usepackage{graphicx}
\usepackage{textcomp}
\usepackage{xcolor}
\usepackage{algorithm}
\usepackage{multirow}
\usepackage{makecell}
\usepackage{bbding}
\usepackage{array}
\usepackage{bm}  
\usepackage{float}
\usepackage{stfloats}
\usepackage{balance}
\usepackage{hyperref}

\def\BibTeX{{\rm B\kern-.05em{\sc i\kern-.025em b}\kern-.08em
    T\kern-.1667em\lower.7ex\hbox{E}\kern-.125emX}}
    \DeclareRobustCommand*{\IEEEauthorrefmark}[1]{%
    \raisebox{0pt}[0pt][0pt]{\textsuperscript{\footnotesize\ensuremath{#1}}}}

\begin{document}

\title{Image2Net: Datasets, Benchmark and Hybrid Framework to Convert Analog Circuit Diagrams into Netlists}

\author{
\IEEEauthorblockN{
Haohang Xu\IEEEauthorrefmark{1,3},
Chengjie Liu\IEEEauthorrefmark{1,3},
Qihang Wang\IEEEauthorrefmark{1,3},
Wenhao Huang\IEEEauthorrefmark{1,3},
Yongjian Xu\IEEEauthorrefmark{1,3},
}
\IEEEauthorblockN{
Weiyu Chen\IEEEauthorrefmark{1,3},
Anlan Peng\IEEEauthorrefmark{1,3},
Zhijun Li\IEEEauthorrefmark{3},
Bo Li\IEEEauthorrefmark{3},
Lei Qi\IEEEauthorrefmark{2,3},
Jun Yang\IEEEauthorrefmark{2,3},
Yuan Du\IEEEauthorrefmark{1}, and
Li Du*\IEEEauthorrefmark{1},
}
\IEEEauthorblockA{\IEEEauthorrefmark{1}School of Electronic Science and Engineering, Nanjing University, Nanjing, China}
\IEEEauthorblockA{\IEEEauthorrefmark{2}School of Integrated Circuit, South East University, Nanjing, China}
\IEEEauthorblockA{\IEEEauthorrefmark{3}National Center of Technology
Innovation for EDA, Nanjing, China}
\IEEEauthorblockA{*Corresponding Author: Li Du \quad Email: ldu@nju.edu.cn}
}

\maketitle

\begin{abstract}
Large Language Model (LLM) exhibits great potential in designing of analog integrated circuits (IC) because of its excellence in abstraction and generalization for knowledge. However, further development of LLM-based analog ICs heavily relies on textual description of analog ICs, while existing analog ICs are mostly illustrated in image-based circuit diagrams rather than text-based netlists. Converting circuit diagrams to netlists help LLMs to enrich the knowledge of analog IC. Nevertheless, previously proposed conversion frameworks face challenges in further application because of limited support of image styles and circuit elements. Up to now, it still remains a challenging task to effectively convert complex circuit diagrams into netlists. To this end, this paper constructs and opensources a new dataset with rich styles of circuit diagrams as well as balanced distribution of simple and complex analog ICs. And a hybrid framework, named Image2Net, is proposed for practical conversion from circuit diagrams to netlists. The netlist edit distance (NED) is also introduced to precisely assess the difference between the converted netlists and ground truth. Based on our benchmark, Image2Net achieves 80.77\% successful rate, which is 34.62\%-45.19\% higher than previous works. Specifically, the proposed work shows 0.116 averaged NED, which is 62.1\%-69.6\% lower than state-of-the-arts. Our datasets and benchmark are available at \url{https://github.com/LAD021/ci2n_datasets}.

\begin{figure*}[ht]
  \centering

  \includegraphics[width=0.9\linewidth]{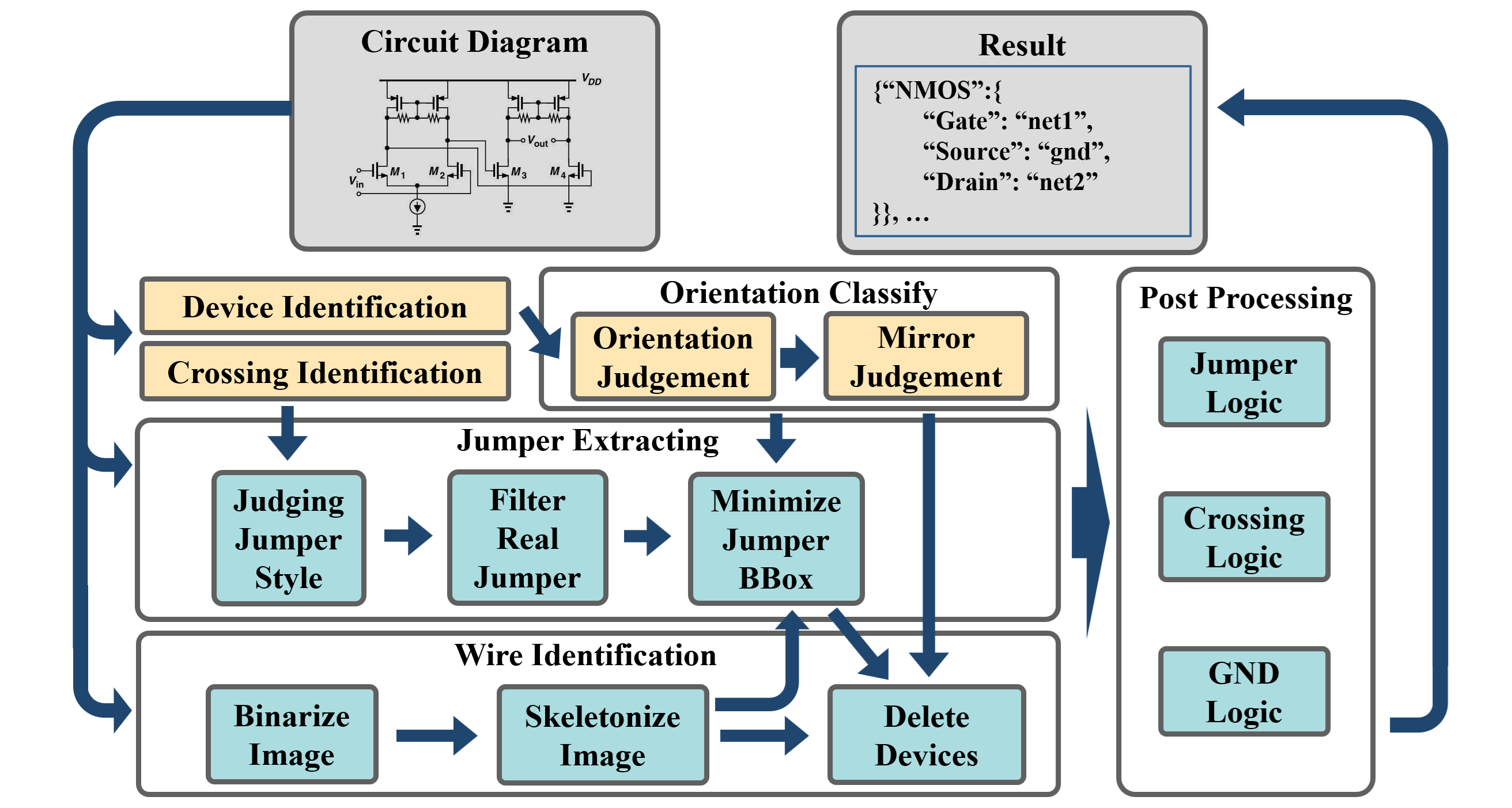}
  \caption{Overall working flow.}
  \label{fig:overall}
\end{figure*}
\end{abstract}


\section{Introduction}
\label{sec:introduction}
Analog Integrated Circuit (IC) design remains challenging due to its heavy reliance on engineer experience and manual effort. While automation could accelerate the process, traditional tools offer limited capabilities due to the complexity of abstracting analog circuits. Large Language Models (LLMs) present new opportunities for converting design specifications into circuit designs through their knowledge understanding and generalization. For example, there are plenty of researches on LLM digital circuit design\cite{Chip-Chat, rtllm, verilogeval, rtlfixer, AutoChip, MEIC}. However, LLMs struggle with generating analog circuit designs due to a severe lack of data, particularly netlist data.

Recent studies like LADAC\cite{liu2024ladac} and AMPAgent\cite{ampagent} explore integrating LLMs with analog design. LADAC uses LLMs to optimize device parameters based on simulation results, while AMPAgent employs a multi-agent framework for automatic Operational Amplifier (OPA) design. Despite these efforts, LLMs still cannot independently generate netlists. Studies such as AMSNet\cite{amsnet} demonstrate that even advanced models like GPT-4 perform poorly in this task. It's largely because analog circuits are often represented as images, not textual netlists, limiting training data for LLMs. More seriously, circuit diagrams are often the core of a textual description of a circuit design method, and the absence of these diagrams can affect the accurate understanding of large amounts of text. This further exacerbates the shortage of high-quality data in analog circuits.

Efforts to convert circuit diagrams to netlists exist but are constrained by varying scales, styles, and complexities of analog ICs. Existing works like AMSNet\cite{amsnet}, CHAI\cite{autospice}, AI-Driven\cite{xiong2024aidriven}, and Img2Sim\cite{Img2Sim-V2} focus on specific, simplified drawing styles, lacking the ability to handle diverse or complex symbols and connections. The quality of training datasets significantly impacts conversion performance, but current datasets, such as those from AMSNet\cite{amsnet} and V3que\cite{v3qwe}, are insufficient for large-scale, complex circuit recognition tasks. Moreover, Existing benchmarks lack practical metrics to evaluate the ability of frameworks to transform complex circuit diagrams. While accuracy is commonly used, it is insufficient as it does not detail the errors in specific cases. CHAI\cite{autospice} uses Graph Edit Distance (GED) for evaluation, but its graph construction from netlists ignores port information, potentially misclassifying incorrect results as correct.

To address the issues mentioned, we open-source our dataset, which includes device identification, crossing identification, and device orientation classification datasets. We also propose a benchmark with manually annotated netlists for evaluation and introduce an evaluation metric called Netlist Edit Distance (NED). Finally, we present our framework, Image2Net, for robustly converting circuit diagrams to netlists. Our datasets and benchmark are available at \cite{mydataset}.

The contributions of this paper are as follows:
\begin{itemize}
\item{\textbf{Diverse Datasets}: We release datasets for device identification, crossing identification, and device orientation classification, featuring rich image styles and complex annotations. These datasets include 2914 circuit diagrams and 84,195 annotations from textbooks, papers, and the internet. They cover comprehensive device types and specialized annotations for crossings and orientations.}
\item{\textbf{Evaluation Metric and Dataset}: We propose an efficient framework for evaluating transitions from circuit diagrams to netlists using the Netlist Edit Distance (NED) metric. NED accurately assesses connection correctness while ignoring irrelevant factors like device or network names. Our evaluation dataset includes 104 manually verified pairs, with more cases being added.}
\item{\textbf{Image2Net Framework}: It utilizes hybrid methods including neural networks and a series of computer vision (CV) algorithms to solve the above problem. The innovation lies in initially determining the overall drawing style through the detection of elements, and then judging the information conveyed by each element in the circuit diagram based on the drawing style and subsequently processing accordingly, thus enabling a practical implementation of multi-style complex circuit recognition. Our framework achieves 80.77\% accuracy, and 0.657 $\overline{NED}$ in our benchmark.}
\end{itemize}

\section{Dataset and Validation Method}
\label{sec:dataset}
The challenges in circuit diagram recognition mainly lie in two aspects: the variety of styles and the existing of interference elements. For the first point, devices and crossings in different circuit diagram follows different drawing rules, see Appendix \ref{sec:a_variety}. Errors in symbol recognition not only lead to type errors in the elements of the netlist, crossing recognition mistakes can directly lead to wrong circuit topology. Secondly, if text, dash-lines or other interfering elements intersect with the bounding boxes used for device recognition, it affects the recognition outcome. We categorize 22 types of devices and 3 types of crossings, define their meanings, provide illustrations, and explain how different crossing styles affect the circuit topology. This content is detailed in Appendix \ref{sec:a_definition}. In order to handle such a variety of image styles, we have established the following datasets and testing methods to assist in training and validation.

\subsection{Datasets}
We made the following datasets open-source: the device identification dataset, the crossing identification dataset, the device orientation classification dataset, and the netlist evaluation dataset. These datasets are used to train the algorithm for each intermediate step in the circuit image-to-netlist process, and for automated evaluation of the final outcome. 

\subsubsection{Device Identification Dataset}
\label{sec:device_dataset}
We open-sourced 2914 complete circuit images, which feature diverse circuit styles and complexities and are derived from papers, textbooks, the Internet, etc. In these images, we annotated the devices according to our specifications and the total number of devices is 48930.
\subsubsection{Crossing Identification Dataset}
\label{sec:crossing_dataset}
We annotate all the crossings of circuits in 1983 images from the aforementioned 2914-image dataset and classify them into three types. The total number of crossings is 28195.

\subsubsection{Device Orientation Dataset}
\label{sec:orientation_dataset}
We collated a dataset for the devices that need orientation and mirror determination. These device images were snippets by results of device object detection. We collected the data of the up, right, down, and left orientations of \texttt{MOS}, \texttt{BJT}, \texttt{Diode}, \texttt{AMP} devices, as well as the mirrored data of the two orientations of \texttt{BJT} and \texttt{AMP} devices. The total number of our annotation is 12086.

\subsubsection{Netlist Dataset}
\label{sec:netlist_dataset}
We provide 104 datasets with labeled netlists as a validation set, provided in JSON format. The circuits in this set ranging from simple to complex.

The detailed statistics is in Appendix \ref{sec:a_statistics}.

\subsection{Evaluation Metric: NED}
\label{sec:ged}
To check the accuracy of our circuit diagram recognition algorithm on the dataset, we proposed a set of consistency checks, and invented our metric called netlist NED. First, we transform the circuit netlist into an equivalent heterogeneous graph. Subsequently, the GED between the identified resulting graph and the golden graph is calculated as the quantitative outcome of the consistency check. We then divide the GED by the total number of devices, nets, and ports in the golden graph. The result becomes NED.

\begin{figure}[ht]
  \centering
  \includegraphics[width=\linewidth]{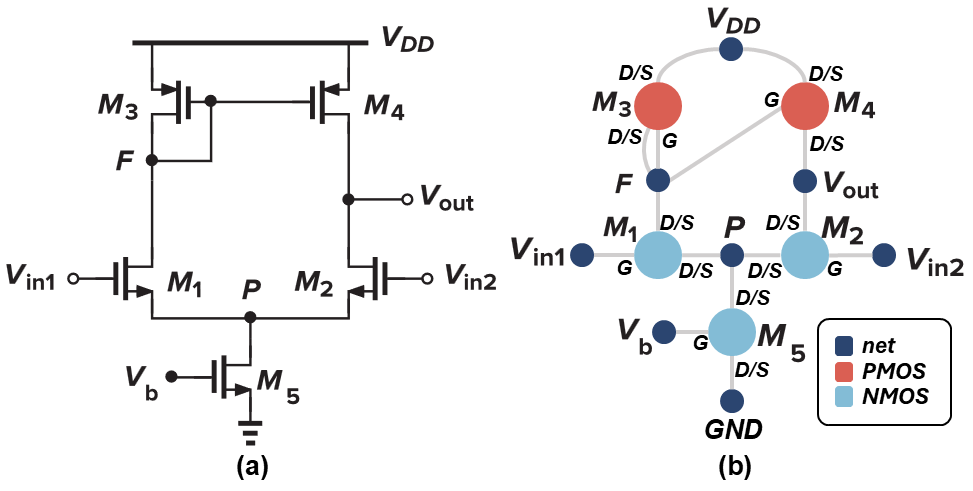}
  \caption{A circuit diagram image and its corresponding graph. The schematic diagram in Fig.(a) encompasses 3 \texttt{NMOS}s, 2 \texttt{PMOS}s, 1 \texttt{Resistor}, and 4 \texttt{NET}s. In Fig.(b), there are altogether 3 types of nodes, namely \texttt{PMOS}, \texttt{NMOS}, \texttt{NET}. There are 4 types of edges, namely \texttt{PMOS\_Gate}, \texttt{PMOS\_D\_S}, \texttt{NMOS\_Gate}, \texttt{NMOS\_D\_S}}
  \label{fig:ged}
\end{figure}

To transform netlist into graph, we define an undirected heterogeneous graph $G = \{V, E\}$, where $V$ and $E$ denote the sets of nodes and edges, respectively. Each node $v \in V$ and each edge $e \in E$ are associated with their mapping functions $\phi(v) \rightarrow A$ and $\phi(e) \rightarrow R$, respectively. $A$ and $R$ denote the sets of node types and edge types, respectively. In the heterogeneous graph defined in this scheme, the elements of the node type set $A$ are various components (such as \texttt{PMOS}, \texttt{NMOS}, \texttt{Capacitor}, \texttt{Resistor}, etc.) and the network (of type \texttt{NET}), and the elements of the edge type set $R$ are the connections from the terminals of each component to the network (such as \texttt{PMOS\_Gate}, \texttt{PMOS\_Drain}, \texttt{PMOS\_Source}, \texttt{Resistor\_Pos}, \texttt{Resistor\_Neg}). Some edges of certain devices are indistinguishable. We internally set their names to be the same and the GED algorithm will consider them to be the same. The indistinguishable ports are listed in Table \ref{table:indistinguishable}.  

\begin{table}
\renewcommand\arraystretch{1.5}
\normalsize
\centering
\caption{Indistinguishable Ports}
\begin{tabular}{|c|c|c|} \hline % 其中，|c|表示文本居中，文本两边有竖直表线。
{Device Type}                & {Ports}         & {Rename}        \\ \hline
MOS                     & Drain, Source       & D\_S    \\ \hline
Source                  & Pos, Neg (Except voltage\_lines)  & Port  \\ \hline
Passive                 & Pos, Neg       & Port              \\ \hline

\end{tabular}
\label{table:indistinguishable}
\end{table}

The computation of the GED is done via an algorithm\cite{abu2015exact}. We invoke the implementation of the NetworkX\cite{networkx} toolkit and perform optimizations to enhance the computational speed. Two nodes or two edges are said to be identical if and only if their types and topological relations are the same. The value of GED is the minimum cost required to perform several insertions, deletions, and substitutions of nodes or edges on a graph G2 to transform it into a graph isomorphic to another graph G1. The cost of all action types is set to 1, and the value of GED is the length of the least action sequence. 

In order to comprehensively evaluate the effects of circuit diagram transformation on netlist, we devised NED. The maximum error count of a circuit can be defined as the total number of devices, nets, and ports of the golden circuit, since this represents the edit distance from an empty graph to a correct circuit. Under normal circumstances, the GED will not exceed this value. We define the NED as the GED calculated by our graph construction method and divided by the maximum error count. Thus the NED result is highly likely to be less than 1. We employ this metric to comprehensively consider the generation effect and the complexity of the problem, providing a normalized and refined indicator for assessing the recognition effect of a graph. Simultaneously, we will take the average NED on the dataset called $\overline{NED}$ as the indicator of the recognition effects of the dataset. We give an example of a netlist with recognition errors and its NED calculation method in Appendix \ref{sec:a_ged}.

In addition, the conventional metric success rate is also considered. The following is the formulation of these evaluation criteria.  

$$
NED = \frac{GED}{N_{device} + N_{net} + N_{port}}
$$

$$
\overline{NED} = \frac{\sum NED_{i}}{N_{total}}
$$

$$
\text{Successful Rate} = \frac{N_{success}}{N_{total}}
$$

Where $N_{success}$ means the number of success cases in a test. $N_{total}$ means the total number of our testcases. $N_{device}$ represents the total number of devices in a netlist annotation.

\subsection{Comparison}

Here is a comparison between our dataset and other open-source datasets. Our dataset is larger in quantity, more complex in circuits, and has more types of annotations. Meanwhile, our dataset is richer in style, derived from books, papers, the web, and encompasses various drawing styles. We also offer a set of validation sets that have undergone manual review.

For other datasets, v3que\cite{v3qwe} provides many images, but the images are not complete circuit diagrams, thus they have only 2.488 devices per image in average. And many of the devices are truncated, which may bring hidden dangers for model training. For AMSNet\cite{amsnet}, their diagram are of similar styles. They provide netlists of each diagram, but they are generated by their solution and they did not make sure that these annotations are manually checked.

The comparison is shown at Table.\ref{table:dataset_compare}

\begin{table}[ht]
\centering
\caption{Comparison with other Datasets}
\begin{tabular}{cccc} % l 表示左对齐的列，这里定义了三列
    \toprule % 顶部粗线
    \textbf{} & \textbf{Image2Net} & \textbf{AMSNet} & \textbf{v3que} \\
    \midrule % 中间线，分隔表头和内容
    Number of Images & \textbf{2914} & 894 & 4761$^{\mathrm{a}}$ \\
    Complete Diagrams & \textbf{yes} & \textbf{yes} & no \\
    Average Devices per Image & \textbf{15.070} & 10.039 & 2.488 \\
    Diagrams Over 20 Devices & \textbf{775} & 36 & NA \\
    Device Annotations & \textbf{yes} & \textbf{yes} & \textbf{yes} \\
    Crossing Annotations & \textbf{yes} & no & no \\
    Orientation Annotations & \textbf{yes} & \textbf{yes} & no \\
    Evaluation Set & \textbf{yes} & no & no \\
    \bottomrule % 底部粗线
    \multicolumn{3}{l}{$^{\mathrm{a}}$v3que provides only randomly sliced images of circuit diagrams, }\\
    \multicolumn{3}{l}{instead of complete circuit diagrams.}
\end{tabular}
\label{table:dataset_compare} % 标签，用于引用
\end{table}

\section{Method}
\label{sec:method}
% \begin{figure*}
%     \centering
%     \includegraphics[scale=0.20]{assets/Method_general_flowchart.png}
%     \caption{{\bf The overall workflow of \PaperName}. It's very good.}
%     \label{fig:method_general_flowchart}
% \end{figure*}

\begin{figure}[ht]
  \centering
  \includegraphics[width=\linewidth]{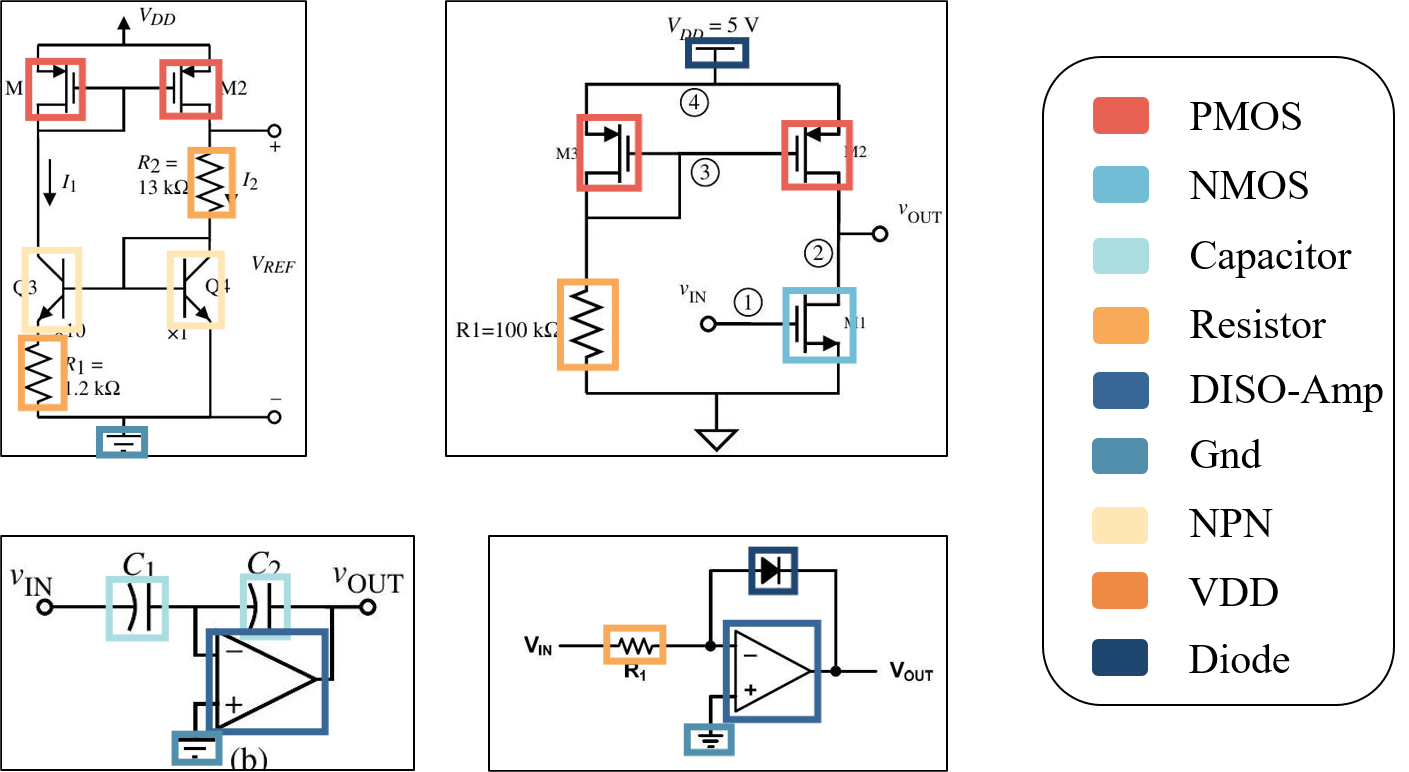}
  \caption{Device identification Examples.}
  \label{fig:device}
\end{figure}

Our overall solution is described in Fig.\ref{fig:overall}. We employ YOLO-V8\cite{yolov8} to identify the positions and types of all components and crossings as in Fig.\ref{fig:device}. We utilize the binary image transferred from the raw image to detect wires. We do a skeleton algorithm and remove small connectivity domains, which eliminates distracting elements such as texts. The port positions are determined by the intersection points between the bounding boxes of the components and the wires, and classification models are used to determine the orientation and mirror information of each component, thereby establishing the port order. The jumper style of the diagram is determined based on the detected crossings, and the crossings considered as jumpers are retained as virtual devices. Finally, the wires connected to each port are identified by employing the geometric overlap method, then the connection information is used to build a netlist. Some post-processing is performed in order to replace virtual components such as jumpers with the correct description of the connection relations in the final generated netlist.

We will provide additional details of some key algorithms below.

\subsection{Wire Identification}
\begin{figure}[!t]
\centering
\subfloat[Raw image]{
		\includegraphics[width=0.20\textwidth]{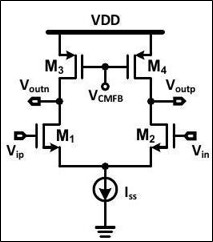}}
\hspace{0.15cm} \subfloat[Skeleton]{
		\includegraphics[width=0.20\textwidth]{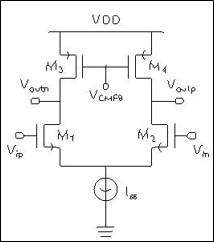}}
\\[0.2cm]
\subfloat[Remove texts]{
		\includegraphics[width=0.20\textwidth]{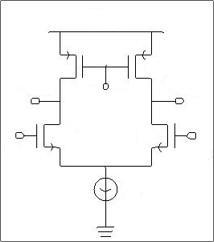}}
\hspace{0.15cm}
\subfloat[Wire detection]{
		\includegraphics[width=0.20\textwidth]{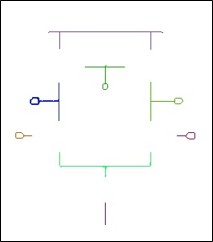}}
\caption{Wire detection algorithm. Fig.(a) is the raw image. Fig.(b) represents the result of the skeleton algorithm. Fig.(c) represents the result of the removing of texts. Fig.(d) shows the final wire detection result.}
\label{fig:wire_detection}
\end{figure}

We eliminate interference during the wire identification step using binarization and skeletonization algorithms. We conduct threshold segmentation on the images and employ the OTSU\cite{OTSU} algorithm for adaptive threshold computation. Subsequently, we skeletonize the binary graph with an algorithm\cite{skeleton_algo}. The width of the wires in the skeletonized binary graph is always $1$. We fill the bounding boxes of all devices with $1$ values and then remove very small connected domains by a threshold of $1/10$ pixels of the largest connected domain from this image to eliminate text and other impurities. We finally remove all the devices in the skeleton image, and then each of the connectivity domain left represents a network. The result is shown in Fig.\ref{fig:wire_detection}.

\subsection{Jumper Identification}
Jumper means an overpass relationship of wires connected to them. The number of wires connected to a jumper must be equal or greater than $4$ and must be even. The intersection of each jumper and its bounding box is the port of the jumper. We can number these ports in clockwise order. Each port is connected to its antipodal in pairs and is not connected to other ports. This relationship can be expressed by the following formula. Let the port number be $i$, the antipodal port number be $j$, the total number of ports be $N$, and $N$ is always an even number:

\[
j = i + \frac{N}{2} \quad \text{when} \quad 1 \leq i \leq \frac{N}{2}
\]

\begin{figure}[ht]
  \centering
  \includegraphics[width=0.8\linewidth]{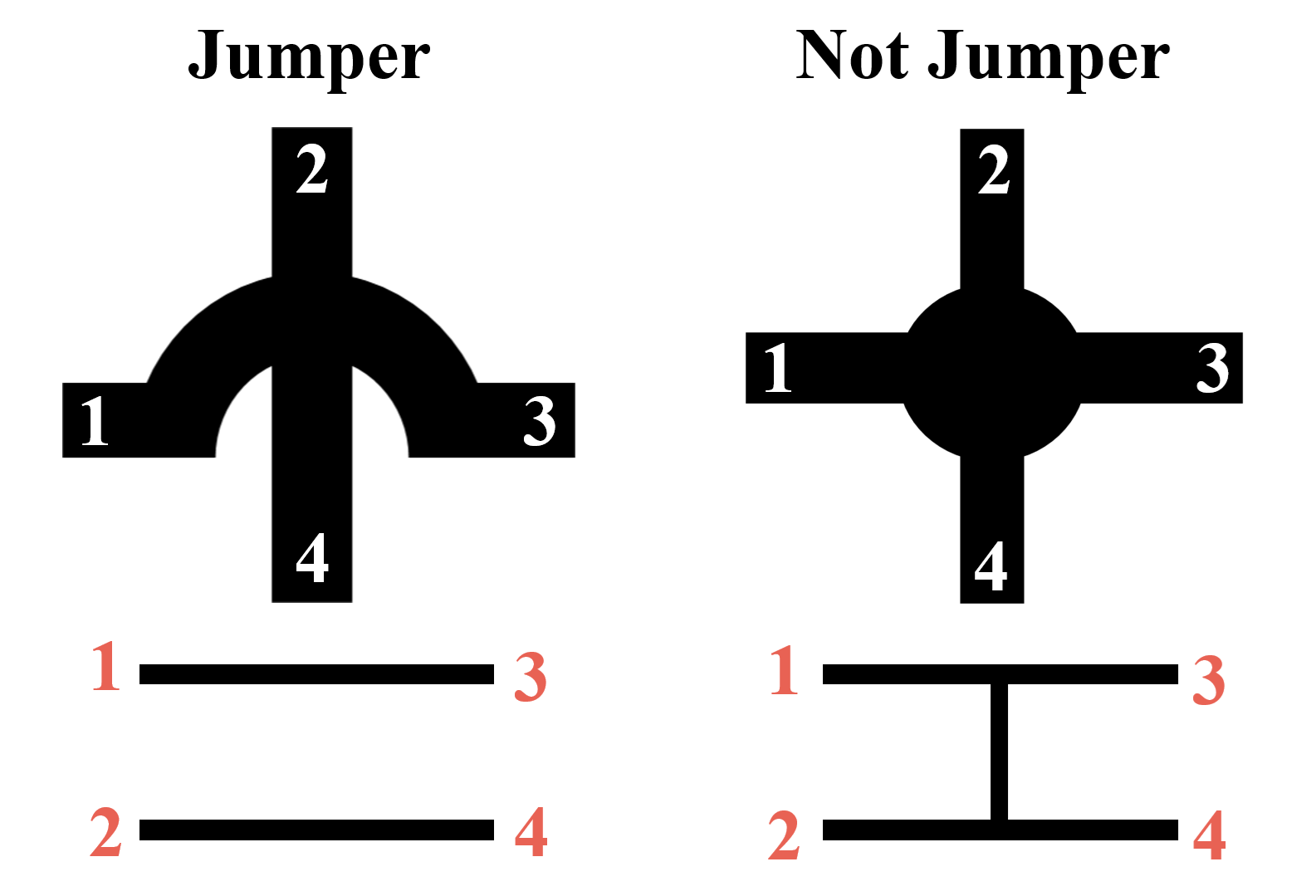}
  \caption{How to pair the ports of a jumper. For a jumper, we name the ports in clockwise order and treat ports in opposite directions as pairs, and the paired ports and their associated nets are connected. For a connection that is not a jumper, all nets connected to it are connected to each other.}
  \label{fig:jumper_connection}
\end{figure}

As shown in Fig.\ref{fig:jumper_connection}, port 1 is connected to port 3, and port 2 is connected to port 4. For non-jumper crossings, all of their ports are connected with each other. We then treat the identified jumpers as devices and also subtract them from the skeleton image.

Following this principle, we adopt some algorithms after using YOLO-V8\cite{yolov8} to identify and filter jumpers. Finally, we treat jumpers we detected as new devices.

\subsection{Port Positioning}
Ports of one device are the intersections of it's bounding box with wires. We use MobileNetV2\cite{mobilenetv2} to identify the orientation and mirror of each device. By these two messages, we can match the intersections one by one with port names, because once you determine a certain port based on the orientation of the device, the remaining ports can always be determined by clock sequence. For some devices have symmetry, we only need to identify the orientation of \texttt{MOS}, \texttt{BJT}, \texttt{AMP}, \texttt{Diode} and \texttt{voltage-lines} by \texttt{Up}, \texttt{Right}, \texttt{Down}, and \texttt{Left}, and mirror of \texttt{BJT} and \texttt{AMP} by \texttt{True} and \texttt{False}.

\subsection{Line Tracking and Post Procedure}
We finally obtain the interconnections of all device ports and output the final netlist. In our wire image, each wire corresponds to a connected domain. We consider all ports that intersect the same connected domain in wire graph to be connected.

\begin{figure}[ht]
  \centering
  \includegraphics[width=\linewidth]{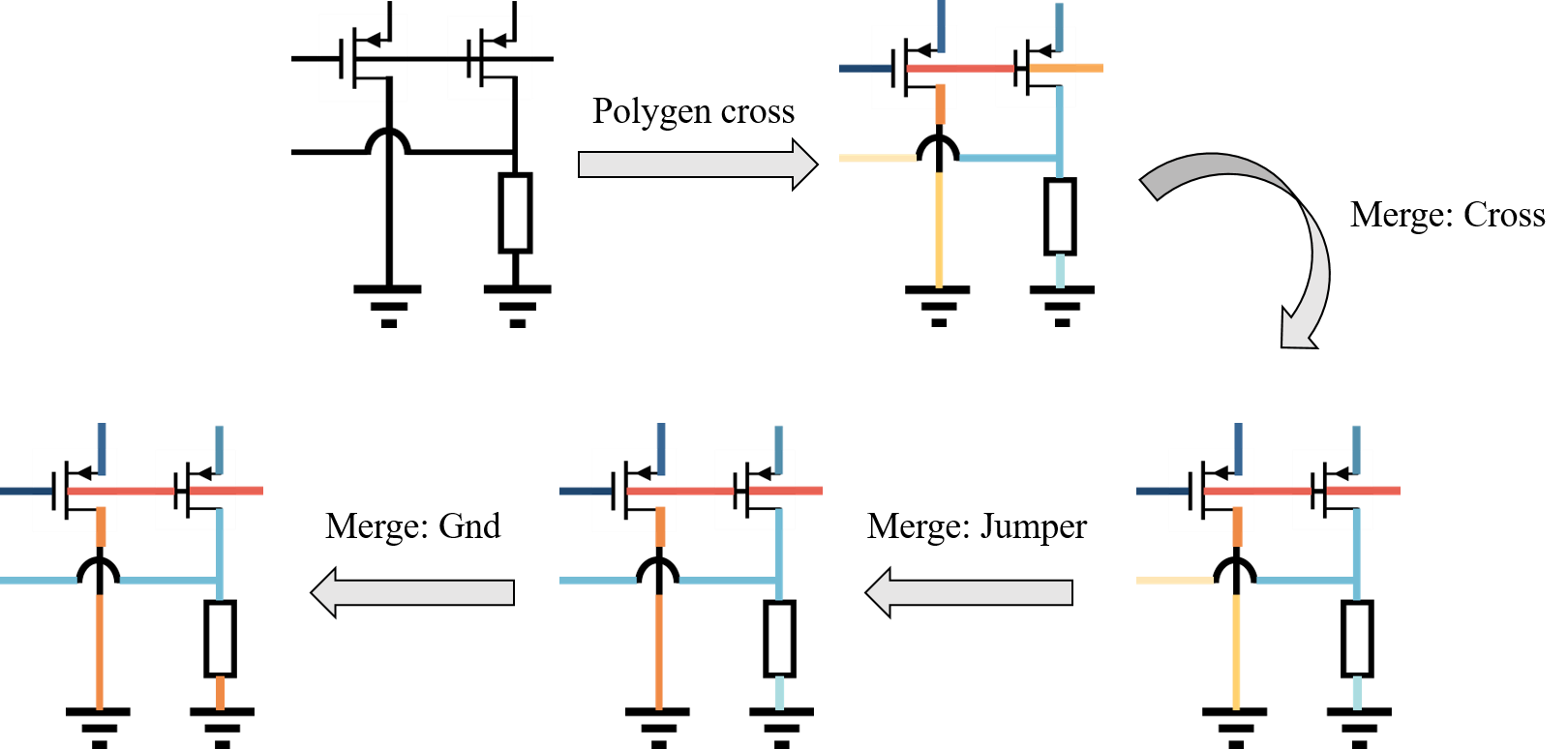}
  \caption{Post processing. Following the working flow, the first figure is the raw image of the circuit diagram. the second figure is the result of connection detection before post processing. In the next steps, the same color means the wires related are all connected. the third to fifth figures shows the merge procedures for cross ports, jumpers, and gnd. }
  \label{fig:wire_patrol}
\end{figure}

Subsequently, we execute some post-processing logic to eliminate some virtual devices and virtual ports. We merged all the networks to which all the \texttt{GND} devices were connected into a single one and removed the \texttt{GND} devices simultaneously. In accordance with the aforementioned jumper logic, we merged the networks connected to the paired ports in the jumper and then deleted the jumper. Finally, we merged the networks connected to the temporary \texttt{Conn} port added to the \texttt{Bulk} device and its \texttt{Gate/Base} port respectively, and then removed this temporary port. These procedures are shown in Fig.\ref{fig:wire_patrol}. Eventually, we obtained the correct netlist.  

\section{Experiment}
\label{sec:experiment}
\subsection{NED Result}
\begin{figure}[ht]  % 每张图片占页面宽度的30%
    \centering
    \includegraphics[width=\linewidth]{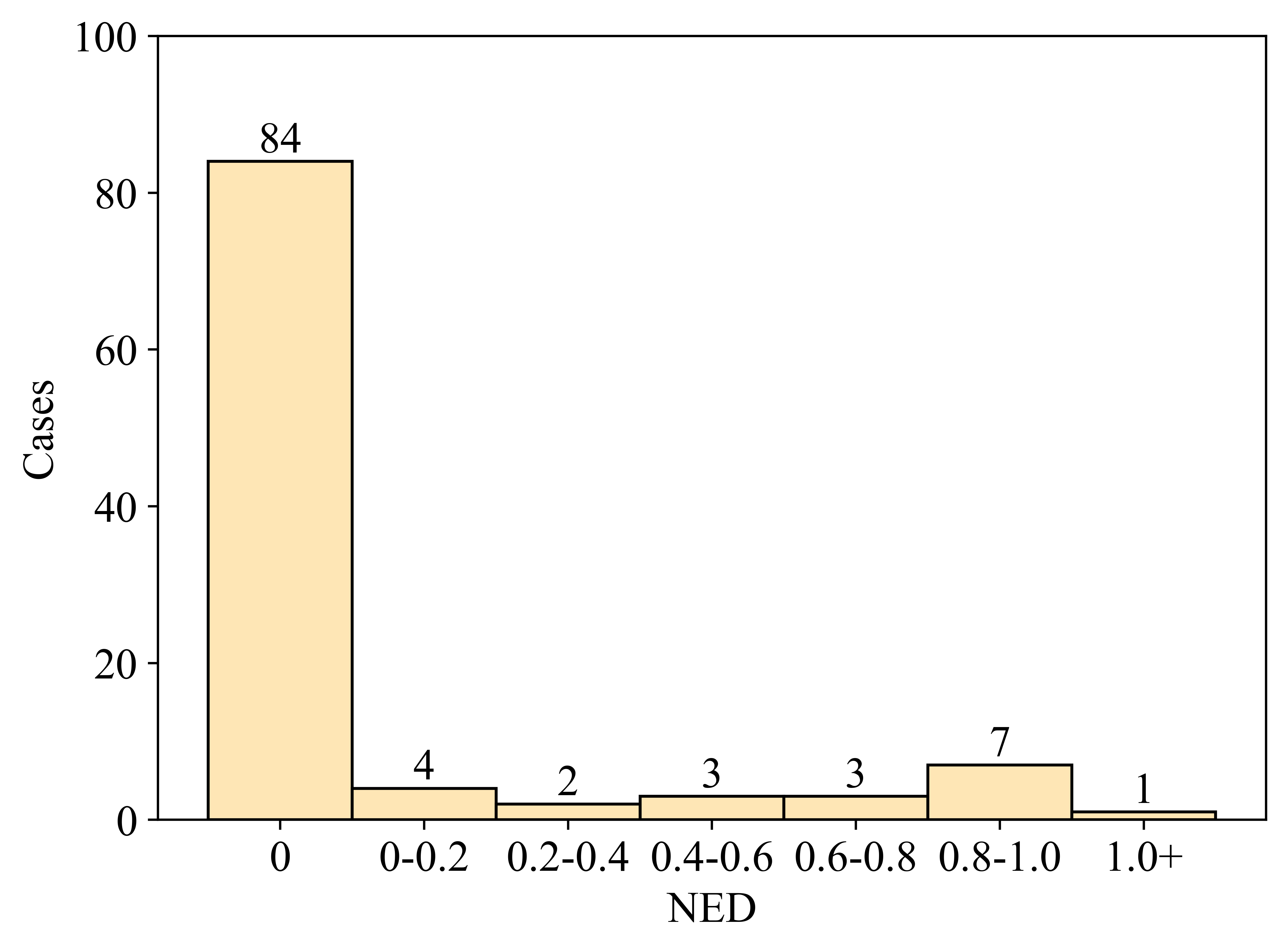}  % 替换为您的图片路径
    \caption{NED distribution. The horizontal axis indicates the NED range, The vertical axis indicates the number of such cases.}
    \label{fig:ged_score}
\end{figure}

To evaluate our algorithm, we generated the netlists using our algorithm on our public test dataset at \ref{sec:netlist_dataset} and calculated the NED in comparison with the golden data. If the result is all correct, the NED should be $0$. After providing each case with sufficient running time, according to our evaluation method in chapter \ref{sec:ged}, we calculated the NED for each case. Our solution ultimately achieved an accuracy rate of 80.77\%, and the $\overline{NED}$ is 0.659. Score distribution is presented in Fig.\ref{fig:ged_score}. According to that the calculation of GED is a NP-Hard problem, we use the algorithm mentioned in \ref{sec:ged} and calculated for 2 hours for each case.

\begin{table}[ht]
\centering
\caption{Comparative Experiment on \\Different Handling Methods of Jumper}
\begin{tabular}{ccc} % l 表示左对齐的列，这里定义了三列
    \toprule % 顶部粗线
    \textbf{Method} & \textbf{Successful Rate} $\uparrow$ & \textbf{$\overline{NED}$ $\downarrow$} \\
    \midrule % 中间线，分隔表头和内容
    Our Method   & \textbf{80.77\%} & \textbf{0.116}  \\
    AMSNet Method   & 35.58\% & 0.306  \\
    CHAI Method & 46.15\% & 0.382  \\
    \bottomrule % 底部粗线
\end{tabular}
\label{table:compares} % 标签，用于引用
\end{table}

\subsection{Comparative Experiment}
We refer to the existing studies. Virtually all studies employ object detection methods to identify devices in images. However, there have been numerous approaches to determine the connection between wires, most of which are rather simplistic. 
Focusing on the judgement of whether a crossing is a jumper, AMSNet\cite{amsnet} makes the assumption that all interleaved wires are not connected, that means crossings are all treated as jumpers. In Masala-
chai\cite{autospice}, jumper logic are not processed and all pixel connections represent true connections by default. These two assumptions and treatment methods are not capable of dealing with complex circuit diagram scenarios. We perform ablation experiments on our framework and simplify the code for handling jumpers based on the above two setups. The results are presented in Table.\ref{table:compares}. These two methods can handle simple images, while complex images with multiple jumper styles are beyond their capabilities. 

Detailed metrics and results of sub-steps in our experiment is shown in Appendix \ref{sec:a_metrics}.

\section{Conclusion}
\label{sec:conclusion}
In this paper, we open-sourced datasets for identifying various elements in the diagrams and designed a GED-based evaluation scheme. Our own solution got 80.77\% accuracy and achieved 0.116 NED on our validation set. Our efforts made it possible to transfer more complicated and varied analog circuit images to netlist, a more concisely and precisely described information form. These data can help LLMs to get a better understanding of analog circuits.

\newpage

% \clearpage
\appendices
\section{Variety of Device and Crossing Types}
\label{sec:a_variety}
\begin{figure}[ht]
  \centering
  \includegraphics[width=\linewidth]{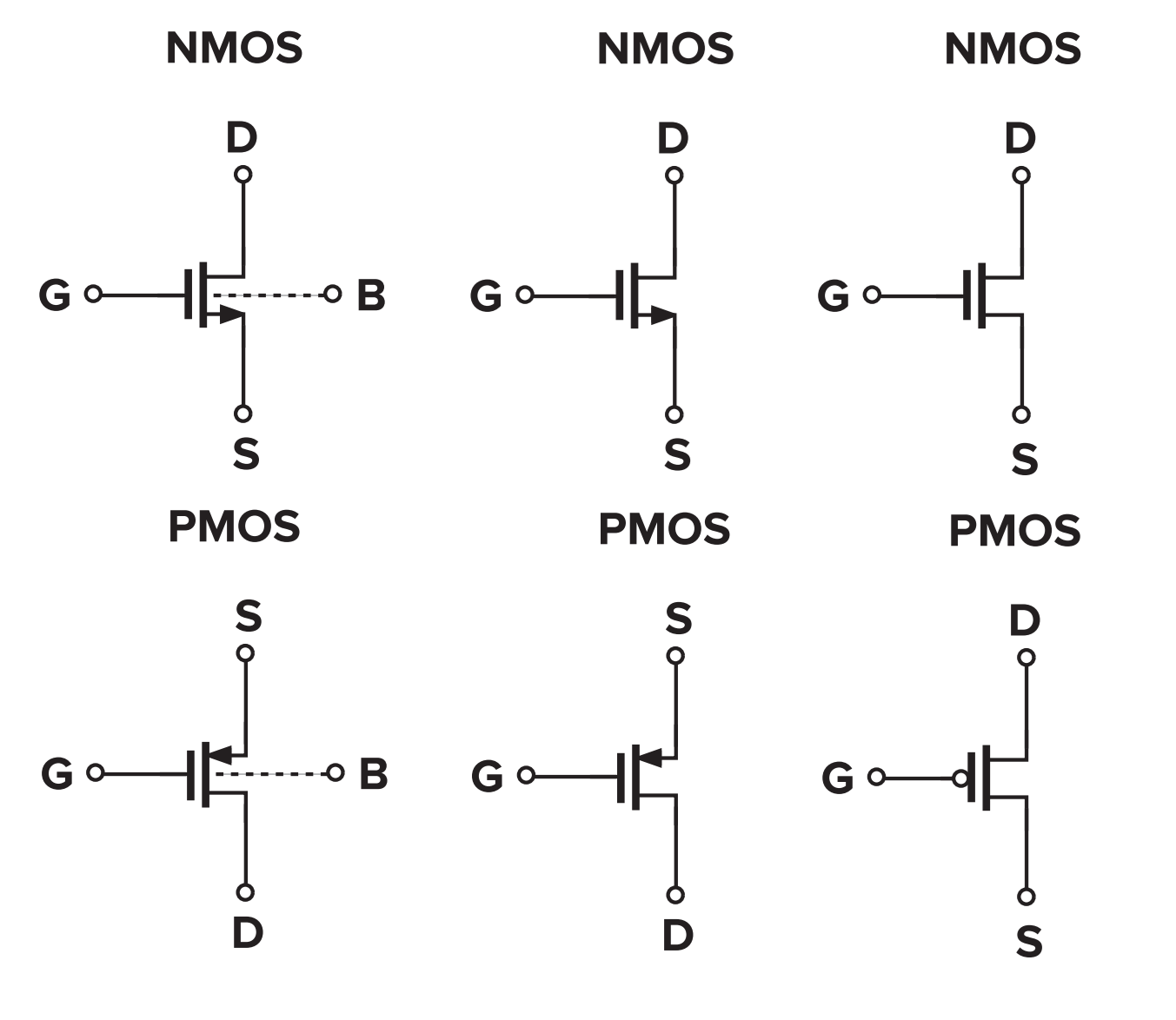}
  \caption{MOS symbols\cite{Razavi2001}.}
  \label{fig:razavi_mos}
\end{figure}

As shown in Fig.\ref{fig:razavi_mos}, for \texttt{MOS} devices only, a variety of drawing styles exist.

\section{Complete definition of our device and crossing types}
\label{sec:a_definition}

\begin{table}
\renewcommand\arraystretch{1.5}
\normalsize
\centering
\caption{Devices and Ports definition}
\begin{tabular}{|c|c|c|c|} \hline % 其中，|c|表示文本居中，文本两边有竖直表线。
Group                & Type in Netlist       & Annotation   & Ports                 \\ \hline
\multirow{6}{*}{MOS} & \multirow{3}{*}{NMOS} & nmos         & \multirow{6}{*}{\makecell{Gate,\\Source,\\Drain,\\(Body)}}  \\ \cline{3-3} 
                     &                       & nmos\_cross  &\\ \cline{3-3} 
                     &                       & nmos\_bulk   &\\ \cline{2-3} 
                     & \multirow{3}{*}{PMOS} & pmos         &\\ \cline{3-3} 
                     &                       & pmos\_cross  &\\ \cline{3-3} 
                     &                       & pmos\_bulk   &\\ \hline
\multirow{4}{*}{BJT} & \multirow{2}{*}{NPN}  & npn          & \multirow{4}{*}{\makecell{Base,\\Collector,\\Emmitter}}  \\ \cline{3-3}
                     &                       & npn\_cross   &\\ \cline{2-3}
                     & \multirow{2}{*}{PNP}  & pnp          &\\ \cline{3-3} 
                     &                       & pnp\_cross   &\\ \hline
\multirow{3}{*}{Amp} & SISO\_Amp              & siso\_amp   & In, Out \\ \cline{2-4}
                     & DISO\_Amp              & diso\_amp   & InP, InN, Out \\ \cline{2-4}
                     & DIDO\_Amp              & dido\_amp   & \makecell{InP, InN,\\OutP, OutN} \\ \hline
Diode                & Diode                  & diode       & \multirow{8}{*}{Pos, Neg} \\ \cline{1-3}
\multirow{3}{*}{Source} & Current                 & current       &  \\ \cline{2-3}
                        & \multirow{2}{*}{Voltage}& voltage       &  \\ \cline{3-3}
                        &                         & voltage\_lines & \\ \cline{1-3}
\multirow{4}{*}{Passive} & \multirow{2}{*}{Resistor} & resistor\_1       &  \\ \cline{3-3}
                         &                           & resistor\_2       &  \\ \cline{2-3}
                         & Capacitor                 & capacitor         &  \\ \cline{2-3}
                         & Inductor                  & inductor          &  \\ \hline
Gnd                  & Gnd                    & gnd         &     undefined   \\ \hline
\end{tabular}
\label{table:list_of_devices}
\end{table}

In this chapter, we will provide our definition of the circuit diagram recognition problem, including our generalization of device symbols and the definition of device ports. 

\begin{figure}[ht]
  \centering
  \includegraphics[width=\linewidth]{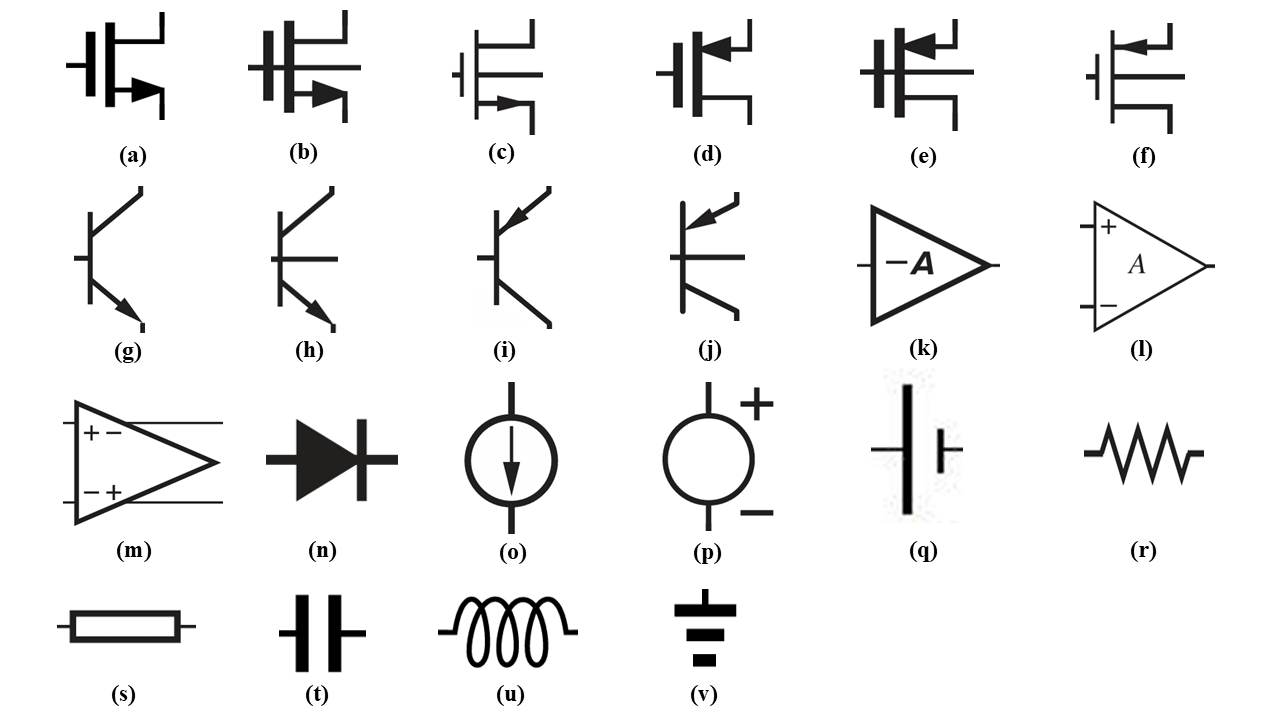}
  \caption{Legends for all devices. \textbf{(a)} nmos \textbf{(b)} nmos\_cross \textbf{(c)} nmos\_bulk \textbf{(d)} pmos \textbf{(e)} pmos\_cross \textbf{(f)} pmos\_bulk \textbf{(g)} npn \textbf{(h)} npn\_cross \textbf{(i)} pnp \textbf{(j)} pnp\_cross \textbf{(k)} siso\_amp \textbf{(l)} diso\_amp \textbf{(m)} dido\_amp \textbf{(n)} diode \textbf{(o)} current \textbf{(p)} voltage \textbf{(q)} voltage\_lines \textbf{(r)} resistor\_1 \textbf{(s)} resistor\_2  \textbf{(t)} capacitor \textbf{(u)} inductor \textbf{(v)} gnd
}

\label{fig:devices_example}
\end{figure}

We categorize 22 types of devices listed in Table \ref{table:list_of_devices}. If there are highly distinct styles within each type, they will be further subdivided into subtypes. The examples of each type are provided in Fig.\ref{fig:devices_example}. Here is a description of each type: \texttt{pmos} and \texttt{nmos} represent MOS devices with no external connections to the body, while \texttt{pmos\_bulk} and \texttt{nmos\_bulk} represent MOS devices with external connections to the body. \texttt{pnp} and \texttt{npn} represent normal BJT. \texttt{pmos\_cross}, \texttt{nmos\_cross}, \texttt{pnp\_cross}, \texttt{npn\_cross} represent a device with a set of parallel Gate ports or Base ports. This type of device has two Gate/Base ports in the legend. \texttt{voltage} represents a voltage source with a circular icon, and \texttt{voltage\_lines} represents a voltage source with a striped icon. \texttt{resistor\_1} represents a resistor of the American standard, and \texttt{resistor\_2} represents a resistor of the European standard. The characteristic differences between the two icons are striking, particularly the high similarity of the American standard resistor to an inductor. Besides, all the other devices can be comprehended through the literal meanings of their labeled names.

\begin{figure}[!t]
\centering
\subfloat[Bridge]{
		\includegraphics[width=0.10\textwidth]{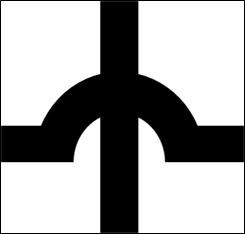}}
\subfloat[Dot]{
		\includegraphics[width=0.10\textwidth]{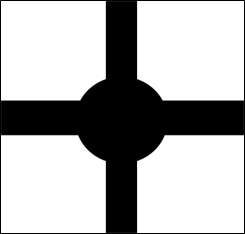}}
\subfloat[Flat]{
		\includegraphics[width=0.10\textwidth]{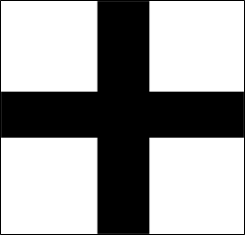}}
\caption{Three types of crossings.}
\label{fig:crossings_example}
\end{figure}

For crossings, as is shown in Fig.\ref{fig:crossings_example}, in general, there are 3 basic styles of symbols for annotating crossings: \texttt{Bridge}, \texttt{Flat}, \texttt{Dot}. Of which types of work as jumpers in a specified diagram, there are several legitimate styles which we have classified:

\begin{itemize}
\item{If any \texttt{Bridge} is found, the \texttt{Bridge} should act as a jumper, and the other symbols of the crossing are to indicate the connection of related wires.}
\item{If there are no \texttt{Bridge} and both \texttt{Dot} and \texttt{Flat} are found, the \texttt{Flat} represents jumper.}
\item{Otherwise, no jumper is found}.
\end{itemize}

\section{Statistics of Datasets}
\label{sec:a_statistics}

\begin{figure*}[!t]  % [t]表示将图放置在页面顶部
    \centering
    \begin{subfigure}{0.31\textwidth}  % 每张图片占页面宽度的30%
        \centering
        \includegraphics[width=\linewidth]{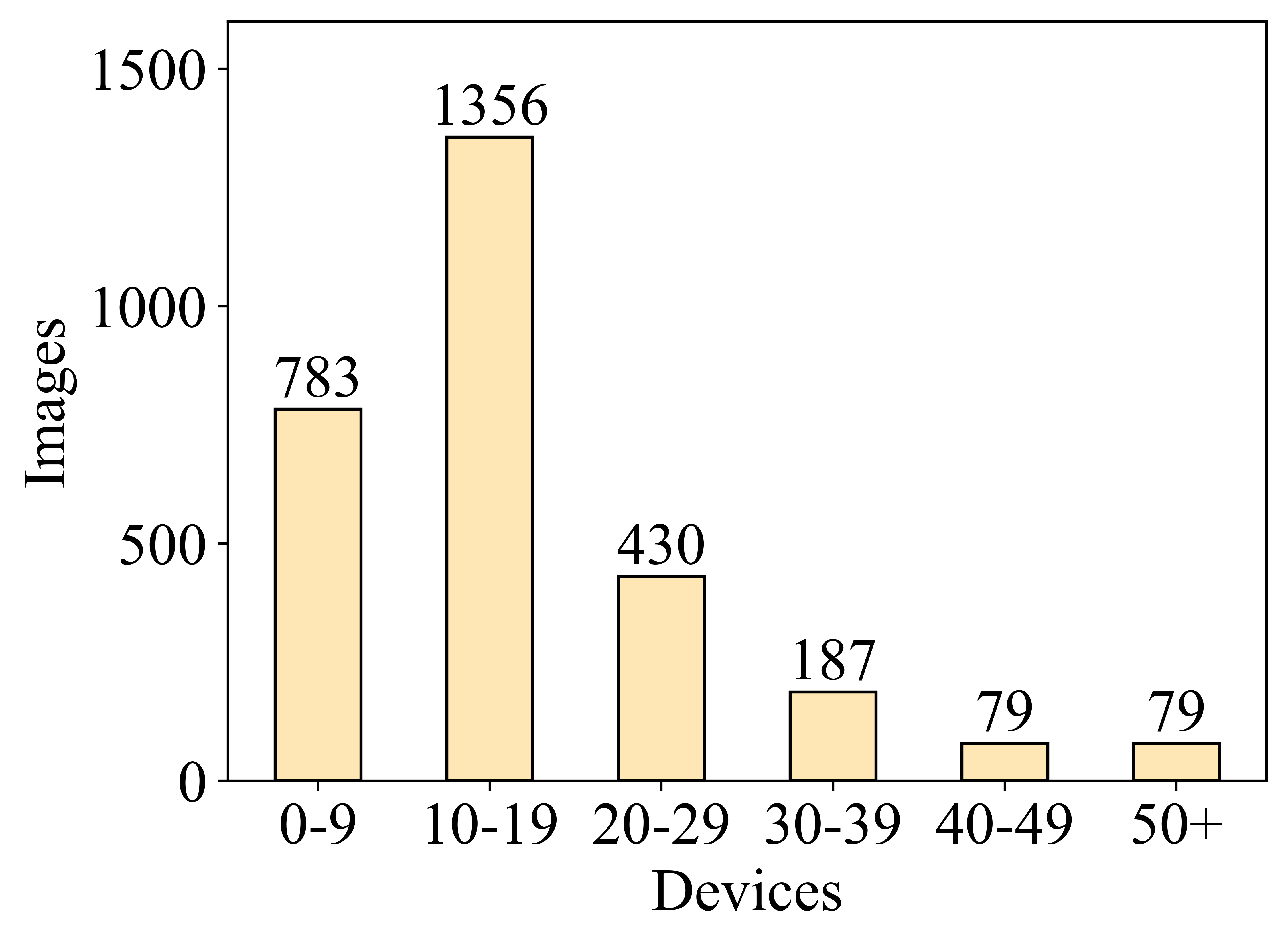}  % 替换为您的图片路径
        \caption{Device number in training dataset}
        \label{fig:component_distribution}
    \end{subfigure}
    \hspace{0.01\textwidth}
    \begin{subfigure}{0.31\textwidth}
        \centering
        \includegraphics[width=\linewidth]{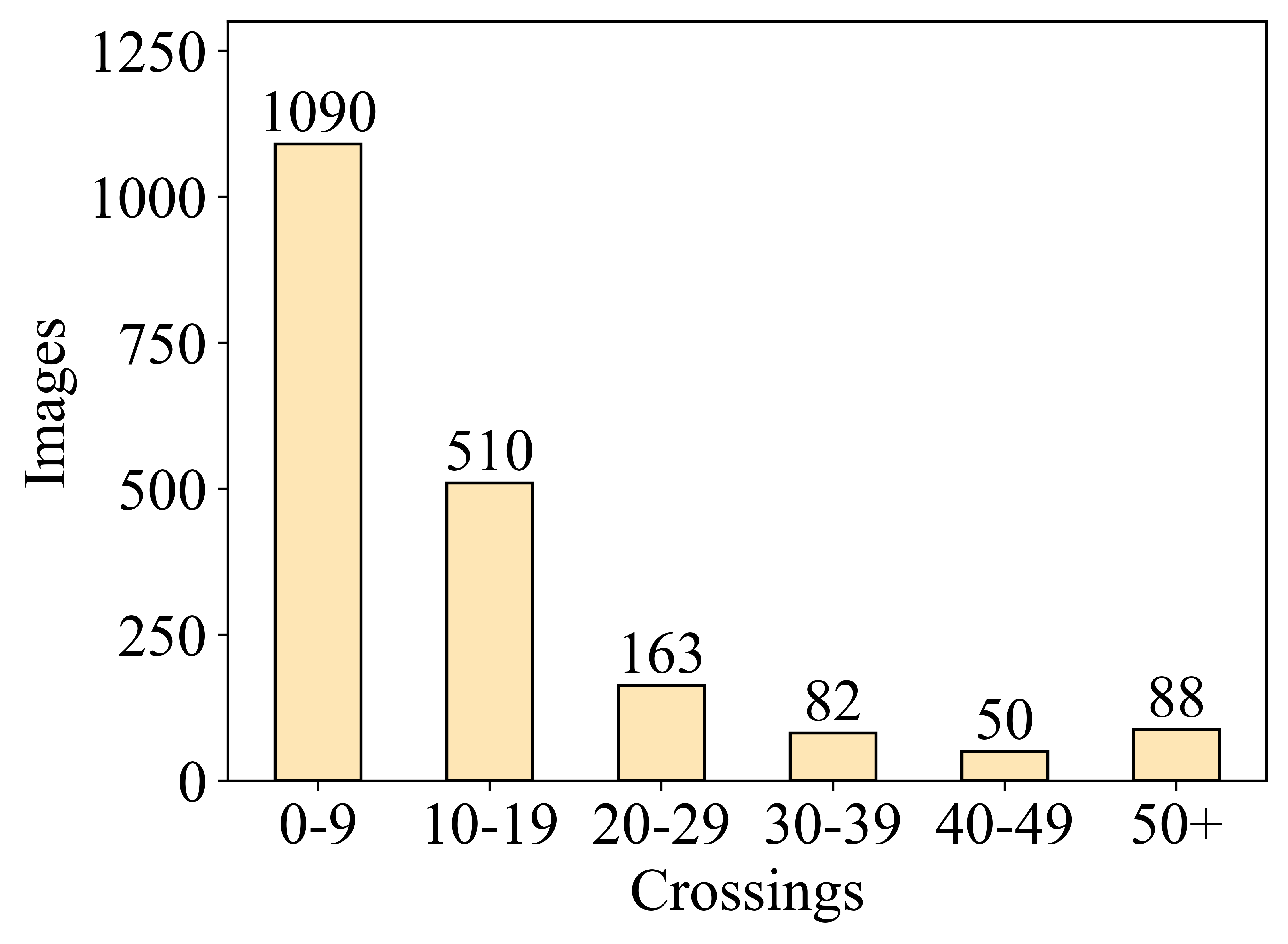}  % 替换为您的图片路径
        \caption{Crossing number in training dataset}
        \label{fig:crossing_distribution}
    \end{subfigure}
    \hspace{0.01\textwidth}
    \begin{subfigure}{0.31\textwidth}
        \centering
        \includegraphics[width=\linewidth]{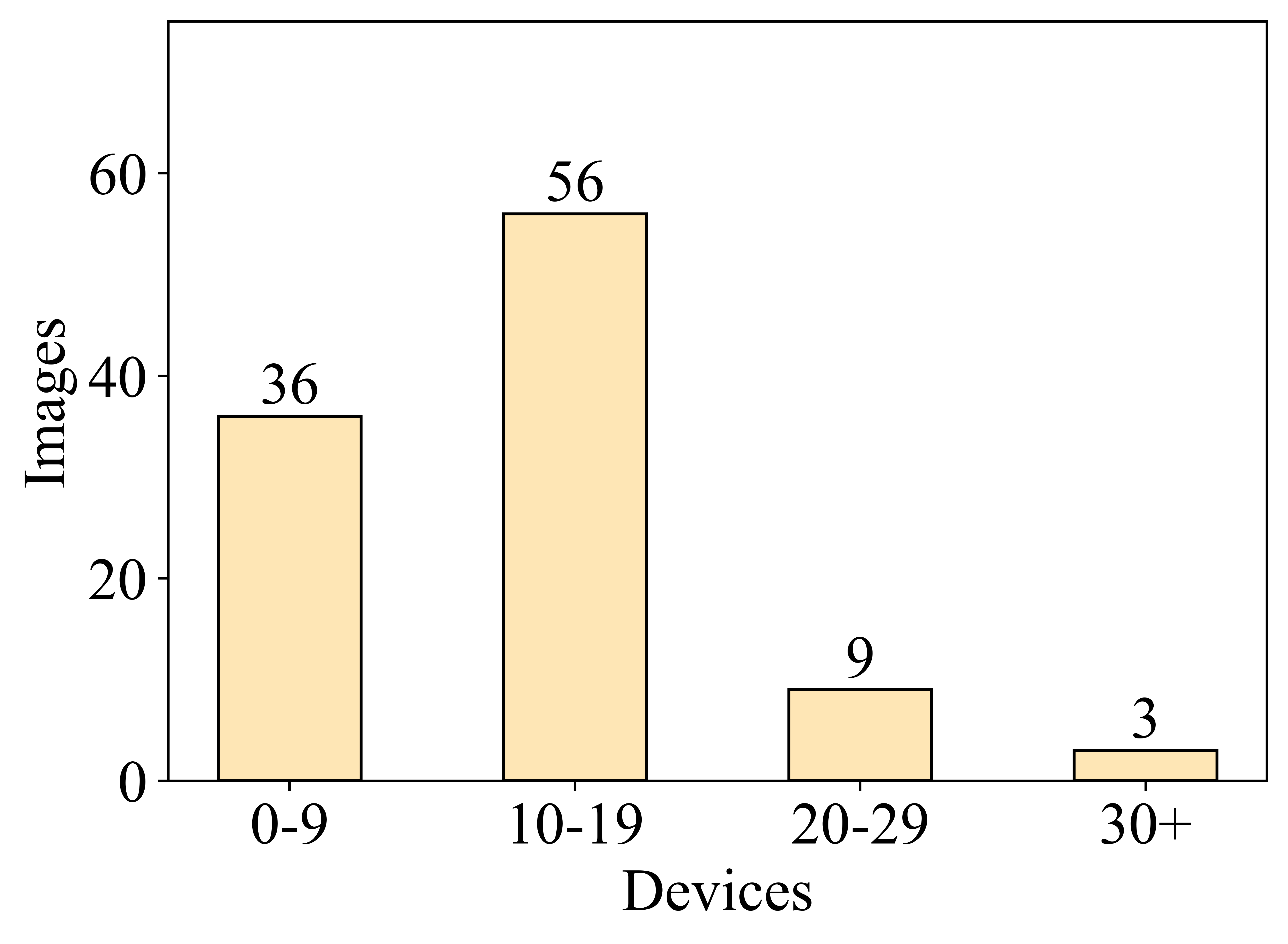}  % 替换为您的图片路径
        \caption{Device number in evaluation dataset}
        \label{fig:netlist_distribution}
    \end{subfigure}
    \caption{Data distribution in our datasets. The horizontal axis indicates the number of devices per image, the vertical axis represents the number of such images. These histograms indicate the complexity of our datasets.}
    \label{fig:data_distribution_all}
\end{figure*}

The quantity distribution of various components and crossings in the device identification dataset (see Chapter \ref{sec:device_dataset}) and crossing identification dataset (see Chapter \ref{sec:crossing_dataset}) is shown in Table.\ref{table:statistics_of_devices}, and the distribution of various components in device orientation dataset (see Chapter \ref{sec:orientation_dataset}) is shown in Table.\ref{table:sum_orient_dataset}. We judge the complexity of the circuit diagram by the number of components and crossings. The distribution of the number of components in the training set images is shown in Fig.\ref{fig:data_distribution_all}(a), and the distribution of crossings is shown in Fig.\ref{fig:data_distribution_all}(b). The distribution of the validation set (see Chapter \ref{sec:netlist_dataset}) is shown in Fig.\ref{fig:data_distribution_all}(c).

\begin{table}
\renewcommand\arraystretch{1.5}
\normalsize
\centering
\caption{Statistics of Devices and Crossings}
\begin{tabular}{|c|c|c|c|} \hline % 其中，|c|表示文本居中，文本两边有竖直表线。
Group                & Total Count       & Annotation   & Count                 \\ \hline
\multirow{6}{*}{MOS} & \multirow{6}{*}{11316} & nmos         &5286
\\ \cline{3-4} 
                     &                       & nmos\_cross  &282
\\ \cline{3-4} 
                     &                       & nmos\_bulk   &670
\\ \cline{3-4} 
                     &                       & pmos         &4138
\\ \cline{3-4} 
                     &                       & pmos\_cross  &477
\\ \cline{3-4} 
                     &                       & pmos\_bulk   &463
\\ \hline
\multirow{4}{*}{BJT} & \multirow{4}{*}{3588}  & npn          &2304
\\ \cline{3-4}
                     &                       & npn\_cross   &41
\\ \cline{3-4}
                     &                       & pnp          &1215
\\ \cline{3-4} 
                     &                       & pnp\_cross   &28
\\ \hline
\multirow{3}{*}{Amp} & \multirow{3}{*}{1044}  & siso\_amp    &154
\\ \cline{3-4}
                     &                       & diso\_amp    &825\\ \cline{3-4}
                     &                       & dido\_amp    &65\\ \hline
Diode                & \multirow{1}{*}{953}  & diode        &953
\\ \cline{1-4}
\multirow{3}{*}{Source} & \multirow{3}{*}{2751} & current       &1507
  \\ \cline{3-4}
                        & & voltage       &1036
  \\ \cline{3-4}
                        &                         & voltage\_lines &208
 \\ \cline{1-4}
\multirow{4}{*}{Passive} & \multirow{4}{*}{16806} & resistor\_1       & 8178
 \\ \cline{3-4}
                         &                           & resistor\_2       & 1775
 \\ \cline{3-4}
                         &                  & capacitor         & 5728
 \\ \cline{3-4}
                         &                   & inductor          & 1125
 \\ \hline
Gnd                  & 7456                    & gnd         &  7456
      \\ \hline
\multirow{3}{*}{Crossing}     & \multirow{3}{*}{28195}      & bridge      &    1137    \\ \cline{3-4}
                     &                       & dot    &17548\\ \cline{3-4}
                     &                       & flat    &9510\\ \hline
\end{tabular}
\label{table:statistics_of_devices}
\end{table}

\begin{table}
\renewcommand\arraystretch{1.5}
\normalsize
\centering
\caption{Statistics of Device Orientation and Mirror Datasets}
\begin{tabular}{|c|c|c|} \hline % 其中，|c|表示文本居中，文本两边有竖直表线。
Type                & Labels & Total Image Count               \\ \hline
    MOS\_Orientation   & u, r, d, l   & 1561  \\ \hline
    BJT\_Orientation   & u, r, d, l   & 5240  \\ \hline
    BJT\_Mirror        & T, F         & 1310  \\ \hline
    Diode\_Orientation & u, r, d, l   & 1248  \\ \hline
    AMP\_Orientation   & u, r, d, l   & 1388  \\ \hline
    AMP\_Mirror        & T, F         & 369  \\ \hline
    Voltage\_lines\_Orientation   & u, r, d, l   & 970  \\ \hline
\end{tabular}
\label{table:sum_orient_dataset}
\end{table}

\section{An Example of Calculating NED}
\label{sec:a_ged}

\begin{figure}[ht]
  \centering
  \includegraphics[width=0.7\linewidth]{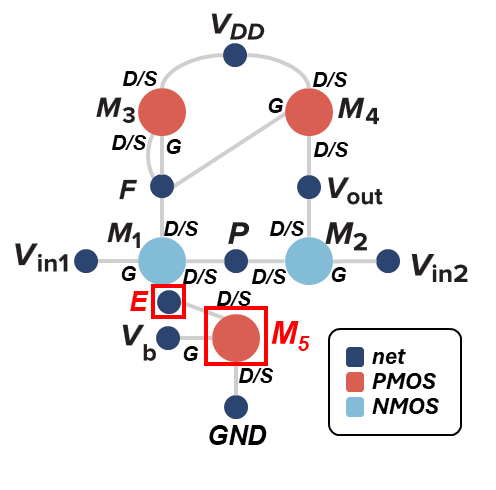}
  \caption{An example of a circuit diagram with a non-zero GED value.}
  \label{fig:ged_detail}
\end{figure}

As illustrated in Fig.\ref{fig:ged_detail}, the two detection errors are indicated with red boxes. The first error is a device type identification error, which mistakenly identifies a \texttt{NMOS} called \texttt{M5} as \texttt{PMOS}; the second error is a port connection error, which incorrectly connects the drain end to the \texttt{E} net instead of the \texttt{P} net. In order to convert the graph $G2$ of the wrong diagram (shown in Fig.\ref{fig:ged_detail}) to the graph $G1$ of the correct diagram (shown in Fig.\ref{fig:ged}(b)), it is necessary to perform 1 edge deletion (delete the edge from \texttt{PMOS} to net \texttt{E}), 1 node deletion (delete the \texttt{E} node) , 1 node substitution (replace \texttt{PMOS} with \texttt{NMOS}), 2 edge substitutions (replace the edges from \texttt{PMOS} to \texttt{Vb} and \texttt{GND} with the edges from \texttt{NMOS} to them) and 1 edge insertion (insert the edge from \texttt{NMOS} to \texttt{P} node). In total, 6 operations are performed, resulting in a GED value of 6. The NED result can be calculated by dividing GED by the sum of total counts of devices, total counts of nets, and total counts of ports. The final result is 0.214.

\section{Detailed Metrics and Results}
\label{sec:a_metrics}

For each model in the process, we showcase their metrics on our dataset. We define the following metrics:

\subsection{Class-Specific Performance}
For each class, we evaluate the results by accuracy and recall. Accuracy means how many classified positive samples are classified correctly, and recall means the number of positive samples that are correctly classified.

\[
\text{Accuracy} = \frac{\text{True Positives}}{\text{True Positives} + \text{False Positives}}
\]

\[
\text{Recall} = \frac{\text{True Positives}}{\text{True Positives} + \text{False Negatives}}
\]

\subsection{Target Detection Performance}
The Mean Average Precision (mAP)@IoU=0.5 (\texttt{mAP@50}) metric measures the average precision across all classes at an Intersection over Union (IoU) threshold of 50\%.
\[
\text{mAP@50} = \frac{1}{N} \sum_{i=1}^{N} \text{AP}_{\text{class}_i}(0.5)
\]
Where:
\begin{itemize}
    \item $N$ is the number of classes.
    \item $\text{AP}_{\text{class}_i}(0.5)$ is the average precision for class $i$ at an IoU threshold of 50\%.
\end{itemize}

The Mean Average Precision (mAP)@IoU=0.5:0.95 (\texttt{mAP@50-95}) metric measures the average precision across all classes using a range of IoU thresholds from 50\% to 95\% in steps of 5\%.
\[
\text{mAP@50-95} = \frac{1}{N} \sum_{i=1}^{N} \left(\frac{1}{10} \sum_{j=0.5}^{0.95} \Delta_j \cdot \text{AP}_{\text{class}_i}(j)\right)
\]
Where:
\begin{itemize}
    \item $N$ is the number of classes.
    \item $\text{AP}_{\text{class}_i}(j)$ is the average precision for class $i$ at an IoU threshold of $j$.
    \item $\Delta_j$ is the step size, which is 0.05 in this case.
\end{itemize}

% \subsection{Sub-step Results}
Here are the detailed results of each step in our framework.

\subsection{Device Identification}
For the device identify result, there is no allowance for the discrimination of the type, and the recognition box should be as precise as possible and fit the edge of the device. Therefore, we adopt \texttt{mAP@50} and \texttt{mAP@50-95}. The result on our dataset at \ref{sec:device_dataset} is \texttt{mAP@50}: 98.38\%, \texttt{mAP@50-95}: 74.31\%.

\subsection{Crossing Identification}
Regarding cross-line recognition, to determine the cross-line style of the image, we identified all the intersections within the image. If some non-cross-line intersections are missed, it might not cause problems, and even some interference such as text being recognized as cross-line will not impact the subsequent line tracking. Also, the recognition box does not need to precisely align with the annotation; it merely requires the point of intersection to be within the recognition box. We devised the following metric to determine the percentage of the points of intersection that fall within the recognition boxes. We define \texttt{mAP@inside}, emulating the definition approach of \texttt{mAP@50}. The condition of average precision is modified from IoU=0.5 to whether the center of the annotation box is within the recognition box.  

\[
\text{mAP@inside} = \frac{1}{N} \sum_{i=1}^{N} \text{AP}_{\text{class}_i}(acip)
\]
Where:
\begin{itemize}
    \item $N$ is the number of classes.
    \item $\text{AP}_{\text{class}_i}(acip)$ is the average precision for class $i$ in the case where the center of the annotation bounding box is inside the prediction bounding box.
\end{itemize}

The result on our dataset at \ref{sec:crossing_dataset} is \texttt{mAP@inside}: 97.30\%.

\subsection{Device Orientation and Mirror Claasification}
Simultaneously, we combine the classifying metric to assess the effects of cross-line recognition. In our dataset, we utilize accuracy and recall to evaluate the effect, and the result of each device on our dataset at \ref{sec:orientation_dataset} can be found in Table \ref{table:result_of_device}.

\begin{table}[ht]
\centering
\caption{Result of Device Orientation and Mirror Detection} % 表格标题
\begin{tabular}{cccc} % l 表示左对齐的列，这里定义了三列
    \toprule % 顶部粗线
    \textbf{Type} & \textbf{Labels} & \textbf{Accuracy} & \textbf{Recall} \\
    \midrule % 中间线，分隔表头和内容
    MOS\_Orientation   & u, r, d, l   & 99.49\% & 99.47\%  \\
    BJT\_Orientation   & u, r, d, l   & 99.90\% & 99.91\%  \\
    BJT\_Mirror        & T, F         & 98.78\% & 98.32\%  \\
    Diode\_Orientation & u, r, d, l   & 99.44\% & 99.47\%  \\
    AMP\_Orientation   & u, r, d, l   & 99.64\% & 99.64\%  \\
    AMP\_Mirror        & T, F         & 95.95\% & 94.74\%  \\
    Voltage\_lines\_Orientation   & u, r, d, l   & 94.85\% & 94.75\% \\
    \bottomrule % 底部粗线
\end{tabular}
\label{table:result_of_device} % 标签，用于引用
\end{table}

\balance
\bibliographystyle{IEEEtran}
\bibliography{ref}

\begin{thebibliography}{10}
\providecommand{\url}[1]{#1}
\csname url@samestyle\endcsname
\providecommand{\newblock}{\relax}
\providecommand{\bibinfo}[2]{#2}
\providecommand{\BIBentrySTDinterwordspacing}{\spaceskip=0pt\relax}
\providecommand{\BIBentryALTinterwordstretchfactor}{4}
\providecommand{\BIBentryALTinterwordspacing}{\spaceskip=\fontdimen2\font plus
\BIBentryALTinterwordstretchfactor\fontdimen3\font minus
  \fontdimen4\font\relax}
\providecommand{\BIBforeignlanguage}[2]{{%
\expandafter\ifx\csname l@#1\endcsname\relax
\typeout{** WARNING: IEEEtran.bst: No hyphenation pattern has been}%
\typeout{** loaded for the language `#1'. Using the pattern for}%
\typeout{** the default language instead.}%
\else
\language=\csname l@#1\endcsname
\fi
#2}}
\providecommand{\BIBdecl}{\relax}
\BIBdecl

\bibitem{Chip-Chat}
\BIBentryALTinterwordspacing
J.~Blocklove, S.~Garg, R.~Karri, and H.~Pearce, ``Chip-chat: Challenges and
  opportunities in conversational hardware design,'' in \emph{2023 ACM/IEEE 5th
  Workshop on Machine Learning for CAD (MLCAD)}.\hskip 1em plus 0.5em minus
  0.4em\relax IEEE, Sep. 2023, p. 1–6. [Online]. Available:
  \url{http://dx.doi.org/10.1109/MLCAD58807.2023.10299874}
\BIBentrySTDinterwordspacing

\bibitem{rtllm}
\BIBentryALTinterwordspacing
Y.~Lu, S.~Liu, Q.~Zhang, and Z.~Xie, ``Rtllm: An open-source benchmark for
  design rtl generation with large language model,'' 2023. [Online]. Available:
  \url{https://arxiv.org/abs/2308.05345}
\BIBentrySTDinterwordspacing

\bibitem{verilogeval}
\BIBentryALTinterwordspacing
M.~Liu, N.~Pinckney, B.~Khailany, and H.~Ren, ``Verilogeval: Evaluating large
  language models for verilog code generation,'' 2023. [Online]. Available:
  \url{https://arxiv.org/abs/2309.07544}
\BIBentrySTDinterwordspacing

\bibitem{rtlfixer}
\BIBentryALTinterwordspacing
Y.-D. Tsai, M.~Liu, and H.~Ren, ``Rtlfixer: Automatically fixing rtl syntax
  errors with large language models,'' 2024. [Online]. Available:
  \url{https://arxiv.org/abs/2311.16543}
\BIBentrySTDinterwordspacing

\bibitem{AutoChip}
\BIBentryALTinterwordspacing
S.~Thakur, J.~Blocklove, H.~Pearce, B.~Tan, S.~Garg, and R.~Karri, ``Autochip:
  Automating hdl generation using llm feedback,'' 2024. [Online]. Available:
  \url{https://arxiv.org/abs/2311.04887}
\BIBentrySTDinterwordspacing

\bibitem{MEIC}
\BIBentryALTinterwordspacing
K.~Xu, J.~Sun, Y.~Hu, X.~Fang, W.~Shan, X.~Wang, and Z.~Jiang, ``Meic:
  Re-thinking rtl debug automation using llms,'' 2024. [Online]. Available:
  \url{https://arxiv.org/abs/2405.06840}
\BIBentrySTDinterwordspacing

\bibitem{liu2024ladac}
C.~Liu, Y.~Liu, Y.~Du \emph{et~al.}, ``Ladac: Large language model-driven
  auto-designer for analog circuits,'' \emph{TechRxiv}, Jan 2024.

\bibitem{ampagent}
\BIBentryALTinterwordspacing
C.~Liu, W.~Chen, A.~Peng, Y.~Du, L.~Du, and J.~Yang, ``Ampagent: An llm-based
  multi-agent system for multi-stage amplifier schematic design from literature
  for process and performance porting,'' 2024. [Online]. Available:
  \url{https://arxiv.org/abs/2409.14739}
\BIBentrySTDinterwordspacing

\bibitem{amsnet}
\BIBentryALTinterwordspacing
Z.~Tao, Y.~Shi, Y.~Huo, R.~Ye, Z.~Li, L.~Huang, C.~Wu, N.~Bai, Z.~Yu, T.-J.
  Lin, and L.~He, ``Amsnet: Netlist dataset for ams circuits,'' 2024. [Online].
  Available: \url{https://arxiv.org/abs/2405.09045}
\BIBentrySTDinterwordspacing

\bibitem{autospice}
\BIBentryALTinterwordspacing
J.~Bhandari, V.~Bhat, Y.~He, S.~Garg, H.~Rahmani, and R.~Karri, ``Masala-chai:
  A large-scale spice netlist dataset for analog circuits by harnessing ai,''
  2024. [Online]. Available: \url{https://arxiv.org/abs/2411.14299}
\BIBentrySTDinterwordspacing

\bibitem{xiong2024aidriven}
W.~Xiong, X.~Meng, Y.~Tao \emph{et~al.}, ``Ai-driven learning and regeneration
  of analog circuit designs from academic papers,'' \emph{Authorea}, Aug 2024.

\bibitem{Img2Sim-V2}
H.~B. Gurbuz, A.~Balta, T.~Dalyan, Y.~D. Gokdel, and E.~Afacan, ``Img2sim-v2: A
  cad tool for user-independent simulation of circuits in image format,'' in
  \emph{2023 19th International Conference on Synthesis, Modeling, Analysis and
  Simulation Methods and Applications to Circuit Design (SMACD)}, 2023, pp.
  1--4.

\bibitem{v3qwe}
\BIBentryALTinterwordspacing
project gtqqq, ``v3qwe dataset,'' \url{
  https://universe.roboflow.com/project-gtqqq/v3qwe }, sep 2022, visited on
  2025-01-03. [Online]. Available:
  \url{https://universe.roboflow.com/project-gtqqq/v3qwe}
\BIBentrySTDinterwordspacing

\bibitem{mydataset}
X.~et~al., ``{ci2n\_datasets},''
  \url{https://github.com/LAD021/ci2n\_datasets}.

\bibitem{abu2015exact}
\BIBentryALTinterwordspacing
Z.~Abu-Aisheh, R.~Raveaux, J.-Y. Ramel, and P.~Martineau, ``An exact graph edit
  distance algorithm for solving pattern recognition problems,'' Lisbon,
  Portugal, pp. 271--278, Jan 2015. [Online]. Available:
  \url{https://hal.archives-ouvertes.fr/hal-01168816}
\BIBentrySTDinterwordspacing

\bibitem{networkx}
\BIBentryALTinterwordspacing
A.~A. Hagberg, D.~A. Schult, P.~J. Swart, and H.~Battley, ``{Exploring network
  structure, dynamics, and function using NetworkX},'' Proceedings of the 7th
  Python in Science Conference, pp. 11--15, 2008. [Online]. Available:
  \url{https://networkx.org/documentation/stable/reference/index.html}
\BIBentrySTDinterwordspacing

\bibitem{yolov8}
Ultralytics, ``{YOLOv8}: Real-time object detection and segmentation,''
  \url{https://yolov8.com/}, 2024, accessed: 2024-10-10.

\bibitem{OTSU}
N.~Otsu, ``A threshold selection method from gray-level histograms,''
  \emph{IEEE Transactions on Systems, Man, and Cybernetics}, vol.~9, no.~1, pp.
  62--66, 1979.

\bibitem{skeleton_algo}
\BIBentryALTinterwordspacing
T.~Y. Zhang and C.~Y. Suen, ``A fast parallel algorithm for thinning digital
  patterns,'' \emph{Commun. ACM}, vol.~27, no.~3, p. 236–239, Mar. 1984.
  [Online]. Available: \url{https://doi.org/10.1145/357994.358023}
\BIBentrySTDinterwordspacing

\bibitem{mobilenetv2}
\BIBentryALTinterwordspacing
M.~Sandler, A.~Howard, M.~Zhu, A.~Zhmoginov, and L.-C. Chen, ``Mobilenetv2:
  Inverted residuals and linear bottlenecks,'' 2019. [Online]. Available:
  \url{https://arxiv.org/abs/1801.04381}
\BIBentrySTDinterwordspacing

\bibitem{Razavi2001}
B.~Razavi, \emph{Design of Analog {CMOS} Integrated Circuits}, 2nd~ed.\hskip
  1em plus 0.5em minus 0.4em\relax McGraw-Hill Education, 2001, figure 2.5 on
  page 10.

\end{thebibliography}

\end{document}